\documentclass[aps,prd,onecolumn,groupedaddress,showpacs,nofootinbib,amssymb, preprint]{revtex4-2}
\usepackage[dvips]{graphicx}
\usepackage{amssymb}
\usepackage{amsmath}
\usepackage{graphicx,,color}
\usepackage{amsfonts}
\usepackage{bm}
\usepackage{cancel}
\usepackage{comment}
\usepackage{floatflt}
\usepackage{slashed}

\newcommand\be{\begin{equation}}
\newcommand\ee{\end{equation}}
\newcommand\nn{\nonumber \\}
\newcommand\e{\mathrm{e}}

\allowdisplaybreaks[4]
\begin{document}

\preprint{KEK-TH-2712, KEK-Cosmo-0379}
\title{
Analogue of Chern-Simons invariant in non-metricity gravity and axion cosmology
}

\author{Shin'ichi~Nojiri$^{1,2}$}\email{nojiri@nagoya-u.jp}
\author{Sergei~D.~Odintsov$^{3,4}$}\email{odintsov@ice.csic.es}

\affiliation{ $^{1)}$ Theory Center, High Energy Accelerator Research Organization (KEK), \\
Oho 1-1, Tsukuba, Ibaraki 305-0801, Japan \\
$^{2)}$ Kobayashi-Maskawa Institute for the Origin of Particles and the Universe, \\
Nagoya University, Nagoya 464-8602, Japan \\
$^{3)}$ Institute of Space Sciences (ICE, CSIC) \\ 
C. Can Magrans s/n, 08193 Barcelona, Spain \\
$^{4)}$ ICREA, Passeig Lluis Companys, 23, 08010 Barcelona, Spain
}

\begin{abstract}

We propose a pseudo-scalar quantity, which is an analogue of the Chern-Simons invariant, in the framework of non-metricity gravity. 
By considering the coupling between the pseudo-scalar quantity and the axion, we give scenarios which may solve the problems 
of the axion misalignment, the $S_8$ problem, and the beginning of inflation. 
When the phase transition associated with the spontaneous breaking of the gauge symmetry of the electroweak theory 
or grand unified theories (GUTs) occurs, the pseudo-scalar quantity has a non-trivial value, which induces the misalignment of the axion field and axion particles are produced. 
If the gradient of the potential is small, the $S_8$ problem might be solved. 
We also propose a mechanism which induces inflation by the misalignment of the axion field generated by the phase transition of the GUTs. 

\end{abstract}

\maketitle

\newpage

\section{Introduction}\label{SecI}

Axion was originally proposed to solve the strong CP problem \cite{Peccei:1977hh}. 
The QCD axion field $\phi$ couples with the instanton density $F\tilde F$, which is a pseudo-scalar and therefore the axion field should also 
be a pseudo-scalar, which has odd parity, that is, under the spatial reflection, the field changes its signature. 
Such a pseudo-scalar field also appears in the string theories as an imaginary counterpart of the dilaton field \cite{Witten:1984dg}. 
The axion field is a natural candidate for dark matter, and it might be produced by the misalignment of the axion field
\cite{Preskill:1982cy, Abbott:1982af, Dine:1982ah}.\footnote{
For pioneer works on axion cosmology, see \cite{Vysotsky:1978dc, Berezhiani:1989fp, Berezhiani:1992rk, Sakharov:1994id, Sakharov:1996xg, Khlopov:1999tm}, 
for examples.
} 

Recently, by Dark Energy Spectroscopic Instrument (DESI) observation \cite{DESI:2024mwx}, 
it has been suggested the possibility that the equation of state parameter $w$ of dark energy had 
a transition from $w<-1$ to $w>-1$ (for recent discussion of observational LCDM-DESI related tensions, see \cite{DiValentino:2025sru}). 
The equation of state (EoS) parameter $w$ is the ratio of the pressure $p$ to the energy density $\rho$, $w=\frac{p}{\rho}$. 
This transition is the inverse of the original ``phantom crossing'' \cite{Hu:2004kh}, where the EoS parameter 
in the early universe is given by $w>-1$ but $w<-1$ today. 
The dark energy with $w<-1$ is called ``phantom''. 
A recent proposal to solve this problem is that, instead of considering the transition from $w<-1$ to $w>-1$ of the dark energy, 
which we may call ``inverse phantom crossing'', the modification of the dark matter sector was considered \cite{Khoury:2025txd}. 
In the scenario, the dark matter decreases more slowly than $1/a^3$, which is usually predicted from the energy conservation of the dust matter ($w=0$). 
Because we are considering the total energy density, the DESI observation seems to indicate that there might have occurred the inverse phantom crossing 
of the dark energy sector only in the case that we assume the usual $1/a^3$ behaviour of the dark matter. 
This scenario may solve the $S_8$ problem \cite{DiValentino:2020vvd}. 
The parameter $S_8$ is defined by matter fluctuation $\sigma_8$ and the matter density parameter $\Omega_m$, $S_8 = \sigma_8 \sqrt{\Omega_m/0.3}$. 
The problem is the discrepancy between the observation of the CMB by assuming the $\Lambda$CDM model and the observation by using lower redshift regions. 

In this paper, we construct a model of an axion field coupled with non-metricity 
gravity~\cite{Nester:1998mp, BeltranJimenez:2018vdo, Runkla:2018xrv, BeltranJimenez:2019tme, Capozziello:2022tvv}. 
By using this theory, we propose models which may solve the problem of the axion misalignment, the $S_8$ problem, and also the beginning of inflation. 

For this purpose, we consider a pseudo-scalar quantity $R_\mathcal{Q}$, which could correspond to the Chern-Simons invariant in Einstein's gravity or 
the instanton density $F\tilde F$. 
The structure of the pseudo-scalar quantity $R_\mathcal{Q}$ proposed in this paper is much simpler than that of the Chern-Simons invariant. 
We mainly work in the model where the action includes a term linear in a non-metricity scalar $Q$ besides the pseudo-scalar term. 
Such a model is, of course, not an $f(Q)$ gravity. 
The reason is that we can choose the coincident gauge, which makes calculations clear. 
We may consider the model where $Q$ in the action is replaced with $f(Q)$, however, it is rather difficult to solve the equation with respect to the connections, 
which are degrees of freedom independent of the metric. 

For the overall evolution of the FLRW universe expansion, the pseudo-scalar term cannot contribute, but this term changes the dispersion relations 
of the left- and right-handed modes of the gravitational wave as in the Chern-Simons gravity~\cite{Alexander:2017jmt, Nojiri:2019nar, Nojiri:2020pqr, Odintsov:2020nwm}. 
The axion is a pseudo-scalar by definition. 
In order to say that the field is pseudo-scalar, we need a coupling with a pseudo-scalar quantity such as the Chern-Simons invariant. 
If there is no such coupling, the field is just a scalar field and cannot be an axion in the original particle physics sense. 
In non-metricity gravity, we did not know what the pseudo-scalar quantity corresponding to the Chern-Simons invariant is 
because the non-metricity scalar $Q$ itself is really a scalar quantity and not a pseudo-scalar. 
Therefore, as a first step, we invent it by using non-metricity tensors. 

Then the next step is to check if the invented quantity $R_\mathcal{Q}$ could really work, and we start with the model 
of the symmetric teleparallel equivalent to general relativity.
The model is the model in non-metricity gravity due to the existence of the pseudo-scalar quantity $R_\mathcal{Q}$ made of the non-metricity tensors. 
As is well-understood, the axion field does not contribute to the time evolution of the FLRW universe as long as the axion field 
is consistently an axion field, that is, the axion field is a pseudo-scalar field with parity odd. 
This property gives strong constraints on the model. 
In the consistent model, the potential, etc., must be an even function of the axion field, and the coupling with the pseudo-scalar quantity like the Chern-Simons invariant, 
the instanton density, or the quantity $R_\mathcal{Q}$ proposed in this paper, must be an odd function of the axion field. 
If not, the parity is explicitly broken, and the axion field cannot be consistently defined to be an axion field. 
If the pseudo-scalar quantity gives a non-trivial contribution to the time evolution of the FLRW spacetime, 
which is parity even, the model is inconsistent. 
If the model is consistent, the pseudo-scalar term can contribute to the fluctuation of the universe, like large-scale structure and also gravitational waves 
and there, the spontaneous parity symmetry breakdown is observed. 
As known in the observation, the parity violation in the large-scale structure has not been observed, and therefore the coupling must be small \cite{Cahn:2021ltp}. 

We choose the potential of the axion field $\phi$ so that the minimum is given by $\phi=0$. 
When there is a phase transition of the electroweak theory or a phase transition of grand unified theories (GUTs) 
associated with the spontaneous breakdown of the gauge symmetries, the pseudo-scalar quantity $R_\mathcal{Q}$ has a non-trivial value. 
Then the coupling of the axion field $\phi$ with $R_\mathcal{Q}$ generates the misalignment of the axion field, which generates the axion particle production. 
If the gradient of potential is small enough, the decrease of the dark matter could be slower than $1/a^3$, which may solve the $S_8$ problem. 
Furthermore, the GUT phase transition may induce inflation. 

In the next section, we briefly review the non-metricity gravity. 
In Section~\ref{SecIII}, we propose the pseudo-scalar quantity made of the non-metricity tensor. 
In Section~\ref{SecIV}, we give a simple model where one can choose the coincident gauge. 
In Section~\ref{SecV}, we consider the FLRW cosmology by using the model in Section~\ref{SecIV}. 
We show that an arbitrary evolution of the FLRW gravity can be realised in the framework of this model. 
In Section~\ref{SecVI}, we investigate the gravitational wave in the model, and we show that the dispersion relations are different and depend on the polarisation. 
In Section~\ref{SecVII}, the mechanism for the misalignment of the axion field and the production of the axion particles are explained. 
In Section~\ref{SecVIII}, we give a scenario to solve the $S_8$ problem. 
In Section~\ref{SecIX}, we discuss the generalisation of the model in Section~\ref{SecIV}. 
The last section is devoted to the summary and conclusion. 

\section{Brief review of non-metricity gravity}\label{SecII}

We express general affine connections on a manifold which is parallelisable and differentiable in the following form, 
\begin{align}
\label{affine}
{\Gamma^\sigma}_{\mu \nu}= {{\tilde \Gamma}^\sigma}_{\mu \nu} + K^\sigma_{\;\mu \nu} + L^\sigma_{\;\mu \nu}\,.
\end{align}
In (\ref{affine}), ${\tilde \Gamma^\sigma}_{\mu \nu}$ expresses the Levi-Civita connection given by the metric 
as in general relativity,
\begin{align}
\label{Levi-Civita}
{{\tilde\Gamma}^\sigma}_{\mu \nu} = \frac{1}{2} g^{\sigma \rho} \left( \partial_\mu g_{\rho \nu} + \partial_\nu g_{\rho \mu}- \partial_\rho g_{\mu \nu}\right)\, ,
\end{align}
${K^\sigma}_{\mu \nu}$ is called contortion, which is defined by using the torsion tensor
${T^\sigma}_{\mu \nu}={\Gamma^\sigma}_{\mu \nu} - {\Gamma^\sigma}_{\nu \mu}$ as follows, 
\begin{align}
\label{contortion}
{K^\sigma}_{\mu \nu}= \frac{1}{2} \left( {T^\sigma}_{\mu \nu} + T^{\ \sigma}_{\mu\ \nu} + T^{\ \sigma}_{\nu\ \mu} \right) \, ,
\end{align}
and, ${L^\sigma}_{\mu \nu}$ represents deformation and is given by
\begin{align}
\label{deformation}
{L^\sigma}_{\mu \nu}= \frac{1}{2} \left( Q^\sigma_{\;\mu \nu} - Q^{\ \sigma}_{\mu\ \nu} - Q^{\ \sigma}_{\nu\ \mu} \right)\,.
\end{align}
In (\ref{deformation}), ${Q^\sigma}_{\mu \nu}$ is the non-metricity tensor defined as,
\begin{align}
\label{non-metricity}
Q_{\sigma \mu \nu}= \nabla_\sigma g_{\mu \nu}= \partial_\sigma g_{\mu \nu} - {\Gamma^\rho}_{\sigma \mu } g_{\nu \rho} 
 - {\Gamma^\rho}_{\sigma \nu } g_{\mu \rho } \,.
\end{align}
In the following, we use $Q_{\sigma\mu\nu}$ to construct the non-metricity gravity. 

We now consider the case without torsion, therefore, we assume ${\Gamma^\sigma}_{\mu\nu} = {\Gamma^\sigma}_{\nu\mu}$. 
Symmetric teleparallel theories of gravity are obtained by requiring the Riemann tensor to vanish, 
\begin{align}
\label{curvatures}
\tilde R^\lambda_{\ \mu\rho\nu} \equiv \Gamma^\lambda_{\mu\nu,\rho} -\Gamma^\lambda_{\mu\rho,\nu} + \Gamma^\eta_{\mu\nu}\Gamma^\lambda_{\rho\eta}
 - \Gamma^\eta_{\mu\rho}\Gamma^\lambda_{\nu\eta} =0 \, .
\end{align}
The solution of (\ref{curvatures}) is written in terms of four scalar fields $\xi^a$ $\left( a = 0,1,2,3 \right)$ 
as follows~\cite{Blixt:2023kyr, BeltranJimenez:2022azb, Adak:2018vzk, Tomonari:2023wcs}, 
\begin{align}
\label{G1B}
{\Gamma^\rho}_{\mu\nu}=\frac{\partial x^\rho}{\partial \xi^a} \partial_\mu \partial_\nu \xi^a \, .
\end{align}
We stress that $\xi^a$'s should be scalar fields and $e^a_\mu\equiv \partial_\mu \xi^a$'s could be identified with vierbein fields. 
Because the system has an invariance under general coordinate transformation, we can often choose the gauge condition ${\Gamma^\rho}_{\mu\nu}=0$, 
which is called the coincident gauge and can be realised by the following choice of $\xi^a$'s, 
\begin{align}
\label{Cgauge}
\xi^a=x^a \, . 
\end{align}
The gauge condition, however, often contradicts the standard metric choices of the FLRW universe 
and the spherically symmetric spacetime. 

Under the infinitesimal transformation, $\xi^a \to \xi^a + \delta \xi^a$, we find, 
\begin{align}
\label{G1}
\Gamma^\rho_{\mu\nu} \to \Gamma^\rho_{\mu\nu} + \delta \Gamma^\rho_{\mu\nu}
\equiv \Gamma^\rho_{\mu\nu} - \frac{\partial x^\rho}{\partial \xi^a} \partial_\sigma \delta\xi^a\frac{\partial x^\sigma}{\partial \xi^b}\partial_\mu \partial_\nu \xi^b
+ \frac{\partial x^\rho}{\partial \xi^a} \partial_\mu \partial_\nu \delta \xi^a \, , 
\end{align}
which is used later to find the field equations given by the variation of the action. 
Since $\xi^a$'s are scalar fields, by the coordinate transformation $x^\mu\to x^\mu + \epsilon^\mu$ with infinitesimally small functions $\epsilon^\mu$, 
the variation of $\xi^a$ is given by $\delta\xi^a = \epsilon^\mu \partial_\mu \xi^a$ and we find 
\begin{align}
\label{G1GCT}
\delta \Gamma^\rho_{\mu\nu}
= \epsilon^\sigma \partial_\sigma \Gamma^\rho_{\mu\nu} - \partial_\sigma \epsilon^\rho \Gamma^\sigma_{\mu\nu} 
+ \partial_\mu \epsilon^\eta \Gamma^\rho_{\eta\nu} + \partial_\nu \epsilon^\eta \Gamma^\rho_{\mu\eta} 
+ \partial_\mu \partial_\nu \epsilon^\rho \, , 
\end{align}
Note that the last term in (\ref{G1GCT}) is nothing but the inhomogeneous term. 
Therefore, the general covariance of the covariant derivative is guaranteed. 
Then we find that $\xi^a$'s are surely scalar fields. 

For the construction of symmetric teleparallel gravity theory, the non-metricity tensor in (\ref{non-metricity}) and the scalar $Q$ of the non-metricity are used. 
The definition of $Q$ is given by, 
\begin{align}
\label{non-m scalar}
Q\equiv g^{\mu \nu} \left( {L^\alpha}_{\beta \nu}{L^\beta}_{\mu \alpha} - {L^\beta}_{\alpha \beta} {L^\alpha}_{\mu \nu} \right)
 -Q_{\sigma \mu \nu} P^{\sigma \mu \nu} \, .
\end{align}
Here the definition of the tensor $P^{\sigma \mu \nu}$ is given by,
\begin{align}
\label{non-m conjugate}
{P^\sigma}_{\mu \nu} \equiv &\, \frac{1}{4} \left\{ - {Q^\sigma}_{\mu \nu} + Q^{\ \sigma}_{\mu\ \nu} + Q^{\ \sigma}_{\nu\ \mu}
+ Q^\sigma g_{\mu \nu}- \tilde{Q}^\sigma g_{\mu \nu} - \frac{1}{2} \left( {\delta^\sigma}_\mu Q_\nu + {\delta^\sigma}_\nu Q_\mu \right) \right\}\, .
\end{align}
In addition, $Q_\sigma$ and $\tilde{Q}_\sigma$ are defined by 
$Q_\sigma \equiv Q^{\ \mu}_{\sigma\ \mu}$ and $\tilde{Q}_\sigma=Q^\mu_{\ \sigma \mu}$. 
By using the above definitions, we obtain, 
\begin{align}
\label{Q}
Q=&\, - \frac{1}{4} g^{\alpha\mu} g^{\beta\nu} g^{\gamma\rho} \nabla_\alpha g_{\beta\gamma} \nabla_\mu g_{\nu\rho}
+ \frac{1}{2} g^{\alpha\mu} g^{\beta\nu} g^{\gamma\rho} \nabla_\alpha g_{\beta\gamma} \nabla_\rho g_{\nu\mu}
+ \frac{1}{4} g^{\alpha\mu} g^{\beta\gamma} g^{\nu\rho} \nabla_\alpha g_{\beta\gamma} \nabla_\mu g_{\nu\rho} \nonumber \\
&\, - \frac{1}{2} g^{\alpha\mu} g^{\beta\gamma} g^{\nu\rho} \nabla_\alpha g_{\beta\gamma} \nabla_\nu g_{\mu\rho} \, .
\end{align}
We should note that the difference between $Q$ and the scalar curvature $\tilde R$ of Einstein's gravity is a total derivative, 
$\tilde R = Q - {\tilde\nabla}_\alpha\left(Q^{\alpha}-\tilde{Q}^{\alpha}\right)$. 
We used a covariant derivative ${\tilde\nabla}_\alpha$, which is defined by using the Levi-Civita connection (\ref{Levi-Civita}). 
This tells that the action linear to $Q$ is equivalent to the Einstein-Hilbert action in Einstein's gravity. 

The action of $f(Q)$ gravity with a function $f(Q)$ of $Q$, 
$S=\int d^4 x \sqrt{-g} f(Q)$, 
is, however, different from the action of $f(R)$ gravity. 
By following the proposition in \cite{Nojiri:2024zab}, we regard the metric $g_{\mu\nu}$ and $\xi^a$ as independent fields hereafter. 

\section{Pseudoscalar quantity}\label{SecIII}

Since the axion field is a pseudo-scalar field, which is parity odd, in order to non-trivially couple with gravity, 
we like to have a pseudo-scalar quantity made of a non-metricity tensor. 
In the case of Einstein's gravity, as a pseudo-scalar quantity, there is a Chern-Simons 
invariant $R_\mathcal{CS} \equiv \tilde\epsilon^{\mu\nu\rho\sigma} g^{\xi\zeta} g^{\eta\tau} \tilde R_{\mu\nu\xi\eta} \tilde R_{\rho\sigma\zeta\tau}$. 
Here $\tilde\epsilon$ is a contravariant tensor given by using the totally antisymmetric Levi-Civita symbols $\epsilon_{\mu\nu\rho\sigma}$, 
\begin{equation}
\label{CS2}
\epsilon_{0123} = - \epsilon^{0123} = 1 \, .
\end{equation}
 From the following property, 
\begin{align}
\label{CS3}
dx^\mu \wedge dx^\nu \wedge dx^\rho \wedge dx^\sigma A_{\mu\nu\rho\sigma}
=& - \frac{1}{4!} dx^\mu \wedge dx^\nu \wedge dx^\rho \wedge dx^\sigma \epsilon_{\mu\nu\rho\sigma}
A_{\mu'\nu'\rho'\sigma'} \epsilon^{\mu'\nu'\rho'\sigma'} \nn
=& d^4 x \sqrt{-g} \frac{1}{\sqrt{-g}}
A_{\mu'\nu'\rho'\sigma'} \epsilon^{\mu'\nu'\rho'\sigma'} \, .
\end{align}
for an arbitrary four-form (covariant) tensor$A_{\mu\nu\rho\sigma}$, 
we find that if we define $\tilde\epsilon^{\mu\nu\rho\sigma}$ by,
\begin{equation}
\label{CS4}
\tilde\epsilon^{\mu\nu\rho\sigma} \equiv
\frac{1}{\sqrt{-g}}\epsilon^{\mu\nu\rho\sigma} \, ,
\end{equation}
one gets that $\tilde\epsilon^{\mu\nu\rho\sigma}$ is surely a contravariant tensor. 
Also, by defining,
\begin{equation}
\label{CS4B}
\epsilon_{\mu\nu\rho\sigma}
= \eta_{\mu\mu'} \eta_{\nu\nu'} \eta_{\rho\rho'} \eta_{\sigma\sigma'}
\epsilon^{\mu'\nu'\rho'\sigma'} \, ,
\end{equation}
we obtain, 
\begin{equation}
\label{CS4C}
\tilde \epsilon_{\mu\nu\rho\sigma}
\equiv g_{\mu\mu'} g_{\nu\nu'} g_{\rho\rho'} g_{\sigma\sigma'}
\tilde\epsilon^{\mu'\nu'\rho'\sigma'}
= \sqrt{-g} \epsilon_{\mu\nu\rho\sigma} \, ,
\end{equation}
which is a covariant tensor. 

By using the non-metricity tensor (\ref{non-metricity}), we now propose the following pseudo-scalar quantity, 
\begin{align}
\label{Qps}
R_\mathcal{Q} \equiv&\, \tilde\epsilon^{\mu\nu\rho\sigma} g^{\xi\zeta} Q_{\mu\nu\xi} Q_{\rho\sigma\zeta} 
= \tilde\epsilon^{\mu\nu\rho\sigma} g^{\xi\zeta} \nabla_\mu g_{\nu\xi} \nabla_\rho g_{\sigma\zeta} \nonumber \\
=&\, \tilde\epsilon^{\mu\nu\rho\sigma} g^{\xi\zeta} \left( \partial_\mu g_{\nu\xi} - \Gamma^\eta_{\mu\xi} g_{\nu\eta} \right) 
\left( \partial_\rho g_{\sigma\zeta} - \Gamma^\tau_{\rho\zeta} g_{\sigma\tau} \right) \, .
\end{align}
Note that $R_\mathcal{Q}$ is not a total differential in general 
because the covariant derivative is not defined by using the Levi-Civita connection. 
Therefore, $R_\mathcal{Q}$ is not a topological density. 

The pseudo-scalar quantity $R_\mathcal{Q}$ is an analogue of the Chern-Simons invariant 
$R_\mathrm{CS} \equiv \tilde\epsilon^{\mu\nu\rho\sigma} R^\tau_{\ \lambda\mu\nu}R^\lambda_{\ \tau\rho\sigma}$ 
because both of $R_\mathcal{Q}$ and $R_\mathrm{CS}$ are pseudo-scalar quantities. 
In some sense, the structure of $R_\mathcal{Q}$ is much simpler than that of $R_\mathrm{CS}$ because the order of the derivative of $R_\mathcal{Q}$ is 
two but that of $R_\mathrm{CS}$ is four. 
In the case of the instanton density, we have $F\tilde F = \epsilon^{\mu\nu\rho\sigma} \mathrm{tr} F_{\mu\nu} F_{\rho\sigma}$, where 
$F_{\mu\nu}$ is defined by $F_{\mu\nu} = \sum_a F_{\mu\nu}^a T_a = \sum_a 
\left\{ \partial_\mu A_\nu^a - \partial_\nu A_\mu^q - f^a_{\ bc}\left( A^b_\mu A_\nu - A^b_\nu A^c_\mu \right)\right\} T_a$. 
Here $A_\mu^a$ is the $a$ component of the gauge field and $T^a$ is the representation of non-abelian algebra, which is a matrix and $f^a_{\ bc}$'s are structure constants 
of the gauge group. 
If we compare Eq.~(\ref{Qps}) with the instanton density, the metric $g_{\mu\nu}$ in (\ref{Qps}) might be identified with the gauge field $A^a_\mu$. 

\section{A simple model}\label{SecIV}

We may consider the following model with axion field $\phi$, 
\begin{align}
\label{model2}
S= \frac{1}{2\kappa^2} \int d^4 x \sqrt{-g} \left\{ f(Q) - \frac{1}{8} \omega\left( \phi^2 \right) \phi^2 \partial_\mu \phi \partial^\mu \phi 
 - V\left( \phi^2 \right) - \alpha\left( \phi^2 \right) \phi R_\mathcal{Q} \right\} \, .
\end{align}
As one will see later, in the model (\ref{model2}), we cannot solve the equation for the connection given by the variation of the action with respect to $\xi^a$ 
even in the spatially flat FLRW spacetime. 
Furthermore, we will consider the case that the spatial part of the universe is neither isometric nor homogeneous, which makes it impossible to find the 
connections satisfying the equation. 

Then we start with a simple model whose action is given by, 
\begin{align}
\label{model1}
S= \frac{1}{2\kappa^2} \int d^4 x \sqrt{-g} \left\{ Q - \frac{1}{8} \omega\left( \phi^2 \right) \phi^2 \partial_\mu \phi \partial^\mu \phi 
 - V\left( \phi^2 \right) - \alpha\left( \phi^2 \right) \phi R_\mathcal{Q} \right\} \, .
\end{align}
Here $\alpha\left( \phi^2 \right)$, 
the coefficient function $\omega\left( \phi^2 \right)$, and the potential $V\left( \phi^2 \right)$ must be a function of $\phi^2$ so that the parity can be well-defined. 
The extra factor $\phi^2$ in the kinetic term is put for later convenience. 
We will find that we can always choose the coincident gauge (\ref{Cgauge}) in the model. 

When $R_\mathcal{Q}$ does not vanish, the effective potential for $\phi$ is given by (\ref{model1}), whose situation is different from that in (\ref{model2}). 
\begin{align}
\label{Veff}
V_\mathrm{eff}\left( \phi^2 \right) = V\left( \phi^2 \right) + \alpha\left( \phi^2 \right) \phi R_\mathcal{Q} \, .
\end{align}
If $V'_\mathrm{eff}\left( \phi^2 \right)=0$ has any non-trivial solution $\left( \phi \neq 0 \right)$, the parity symmetry is broken. 

For the variation of $S_{R_\mathcal{Q}} \equiv - \frac{1}{2\kappa^2} \int d^4 x \sqrt{-g} \alpha \left( \phi^2 \right) \phi R_\mathcal{Q}$ with respect to $\xi^a$, we find 
\begin{align}
\label{vSRQ}
\frac{\delta S_{R_\mathcal{Q}}}{\delta \xi^a}
=&\, 2 \partial_\eta \left[ \alpha \left( \phi^2 \right) \phi \frac{\partial x^\rho}{\partial \xi^a} \frac{\partial x^\eta}{\partial \xi^b}\partial_\mu \partial_\nu \xi^b
\left\{ \epsilon^{\mu\xi\eta\sigma} g^{\nu\zeta} \Gamma^\eta_{\mu\xi} g_{\xi\rho} 
\left( \partial_\eta g_{\sigma\zeta} - \Gamma^\tau_{\rho\zeta} g_{\sigma\tau} \right) 
\right\} \right] \nonumber \\
&\, + 2 \partial_\mu \partial_\nu \left[ \alpha \left( \phi^2 \right) \phi 
\frac{\partial x^\rho}{\partial \xi^a} 
\left\{ \epsilon^{\mu\xi\eta\sigma} g^{\nu\zeta} \Gamma^\eta_{\mu\xi} g_{\xi\rho} 
\left( \partial_\eta g_{\sigma\zeta} - \Gamma^\tau_{\rho\zeta} g_{\sigma\tau} \right) 
\right\} \right] \, .
\end{align}
We should note $\tilde\epsilon^{\mu\nu\rho\sigma} $ is replaced by $\epsilon^{\mu\nu\rho\sigma}$ without tilde $\tilde\ $ due to the existence of 
$\sqrt{-g}$, $\sqrt{-g} \left( \alpha \left( \phi^2 \right) \phi R_\mathcal{Q} \right) 
= \alpha \left( \phi^2 \right) \phi \epsilon^{\mu\nu\rho\sigma} g^{\xi\zeta} Q_{\mu\nu\xi} Q_{\rho\sigma\zeta}$. 
It is clear that in the coincident gauge (\ref{Cgauge}), which gives $\Gamma^\tau_{\rho\zeta}=0$, Eq.~(\ref{vSRQ}) vanishes 
$\frac{\delta S_{R_\mathcal{Q}}}{\delta \xi^a} = 0$. 
In the simple model (\ref{model1}), the $\xi^a$-dependent part of the first term is total derivative because 
$\tilde R = Q - {\tilde\nabla}_\alpha\left(Q^{\alpha}-\tilde{Q}^{\alpha}\right)$. 
Therefore, the equation given by the variation of the action (\ref{model1}) with respect to $\xi^a$ is satisfied by the coincident gauge (\ref{Cgauge}). 
Of course, if we consider the action where $Q$ in (\ref{model1}) is replaced by a function of $Q$, that is, $f(Q)$ as in (\ref{model2}), we cannot always choose the coincident gauge. 

By choosing the coincident gauge (\ref{Cgauge}), because $\tilde R = Q - {\tilde\nabla}_\alpha\left(Q^{\alpha}-\tilde{Q}^{\alpha}\right)$, 
the variation of the action with respect to the metric gives, 
\begin{align}
\label{eq}
&\, \tilde R_{\mu\nu} - \frac{1}{2} g_{\mu\nu} \tilde R = \frac{1}{2} g_{\mu\nu} \left( - \frac{1}{8} \omega\left( \phi^2 \right) \phi^2 \partial_\rho \phi \partial^\rho \phi 
 - V\left( \phi^2 \right) \right) + \frac{1}{8} \omega\left( \phi^2 \right) \phi^2 \partial_\mu \phi \partial_\nu \phi \nonumber \\
&\, + \alpha \left( \phi^2 \right) \phi \tilde\epsilon^{\zeta\xi\rho\sigma} \partial_\zeta g_{\xi\mu} \partial_\rho g_{\sigma\nu} 
+ \alpha \left( \phi^2 \right) \tilde \epsilon^{\xi\zeta\rho\sigma} g_{\mu\zeta} \partial_\xi \left( \phi \partial_\rho g_{\sigma\nu} \right)
+ \alpha \left( \phi^2 \right) \tilde \epsilon^{\xi\zeta\rho\sigma} g_{\nu\zeta} \partial_\xi \left( \phi \partial_\rho g_{\sigma\mu} \right) \, .
\end{align}
In the following, we investigate the cosmology and the gravitational waves. 

\section{FLRW cosmology}\label{SecV}

We now consider the spatially flat FLRW spacetime whose metric is given by 
\begin{align}
\label{FLRW}
ds^2 = - dt^2 + a(t)^2 \sum_{i=1,2,3} \left( dx^i \right)^2 \, .
\end{align}
We also assume that $\phi$ only depends on $t$. 
Then $(t,t)$ and $(i,j)$ components of Eq.~(\ref{eq}) have the following form, 
\begin{align}
\label{F1}
3H^2 =&\, \frac{1}{16} \omega\left( \phi^2 \right) \phi^2 {\dot\phi}^2 + \frac{1}{2} V\left( \phi^2 \right) \, , \\
\label{F2}
- 3H^2 - 2 \dot H=&\, \frac{1}{16} \omega\left( \phi^2 \right) \phi^2 {\dot\phi}^2 - \frac{1}{2} V\left( \phi^2 \right) \, .
\end{align}
Here, the Hubble rate $H$ is defined by $H\equiv \frac{\dot a}{a}$. 
We should note that $ \alpha \left( \phi^2 \right) \phi R_\mathcal{Q}$ term does not contribute to (\ref{F1}) nor (\ref{F2}). 

In (\ref{F1}) and (\ref{F2}), the redefinition of $\phi$ is absorbed into the redefinition of $\omega\left( \phi^2 \right)$ as in 
\cite{Nojiri:2005pu, Capozziello:2005tf}. 
Then one may identify $\phi^2$ with the cosmological time $t$, $\phi^2=t$ and Eqs.~(\ref{F1}) and (\ref{F2}) have the following forms, 
\begin{align}
\label{Fb}
3H^2 = \frac{1}{4} \omega\left( t \right) + \frac{1}{2} V\left( t \right) \, , \quad 
 - 3H^2 - 2 \dot H= \frac{1}{4} \omega\left( t \right) - \frac{1}{2} V\left( t \right) \, ,
\end{align}
which give 
\begin{align}
\label{Fc}
\omega\left( t \right) = - 4 \dot H\, , \quad 
V\left( t \right) = 6H^2 - 2 \dot H \, .
\end{align}
We now assume that the Hubble rate is given by a function $f(t)$, $H=f(t)$. 
Then by choosing 
\begin{align}
\label{Fd}
\omega\left( \phi^2 \right) = - 4 f' \left( \phi^2 \right) \, , \quad 
V\left( \phi^2 \right) = 6f \left( \phi^2 \right)^2 - 2 f' \left( \phi^2 \right) \, .
\end{align}
the universe with $H=f(t)$ can be realised in the model given by (\ref{Fd}). 

\section{Gravitational wave}\label{SecVI}

Let us now consider the propagation of the gravitational wave.
For the following general variation of the metric,
\begin{align}
\label{variation1}
g_{\mu\nu}\to g_{\mu\nu} + h_{\mu\nu}\, ,
\end{align}
and by using (\ref{eq}), we obtain the equation describing the propagation of the gravitational wave as follows,
\begin{align}
\label{gb4bD4B0}
0=&\, \left[ \frac{1}{2} {\tilde R} + \frac{1}{2} \left\{
 - \frac{1}{8} \omega\left( \phi^2 \right) \phi^2 \partial_\mu \phi \partial^\mu \phi 
 - V\left( \phi^2 \right) \right\} \right] h_{\mu\nu}
 - \frac{1}{16} g_{\mu\nu} \omega\left( \phi^2 \right) \phi^2 \partial^\tau \phi \partial^\eta \phi h_{\tau\eta} \nonumber \\
&\, + \frac{1}{2} g_{\mu\nu} \left\{ -h_{\rho\sigma} {\tilde R}^{\rho\sigma} 
+ {\tilde\nabla}^\rho {\tilde\nabla}^\sigma h_{\rho\sigma} - {\tilde\nabla}^2 \left(g^{\rho\sigma}h_{\rho\sigma}\right) \right\} \nonumber \\
&\, - \frac{1}{2} \left\{{\tilde\nabla}_\mu{\tilde\nabla}^\rho h_{\nu\rho}
+ {\tilde\nabla}_\nu {\tilde\nabla}^\rho h_{\mu\rho} - {\tilde\nabla}^2 h_{\mu\nu}
 - {\tilde\nabla}_\mu {\tilde\nabla}_\nu \left(g^{\rho\lambda}h_{\rho\lambda}\right) \right. \nonumber \\
&\, \left. - 2{\tilde R}^{\lambda\ \rho}_{\ \nu\ \mu}h_{\lambda\rho}
+ {\tilde R}^\rho_{\ \mu}h_{\rho\nu} + {\tilde R}^\rho_{\ \mu}h_{\rho\nu} \right\} \nonumber \\
&\, - \frac{\alpha \left( \phi^2 \right)}{2} g^{\tau\eta} h_{\tau\eta} \left\{ \phi \tilde\epsilon^{\zeta\xi\rho\sigma} \partial_\zeta g_{\xi\mu} \partial_\rho g_{\sigma\nu} 
+ \alpha \tilde \epsilon^{\xi\zeta\rho\sigma} g_{\mu\zeta} \partial_\xi \left( \phi \partial_\rho g_{\sigma\nu} \right)
+ \alpha \tilde \epsilon^{\xi\zeta\rho\sigma} g_{\nu\zeta} \partial_\xi \left( \phi \partial_\rho g_{\sigma\mu} \right) \right\} \nonumber \\
&\, + \alpha \left( \phi^2 \right) \phi \tilde\epsilon^{\zeta\xi\rho\sigma} \partial_\zeta h_{\xi\mu} \partial_\rho g_{\sigma\nu} 
+ \alpha \left( \phi^2 \right) \phi \tilde\epsilon^{\zeta\xi\rho\sigma} \partial_\zeta g_{\xi\mu} \partial_\rho h_{\sigma\nu} 
+ \alpha \left( \phi^2 \right) \tilde \epsilon^{\xi\zeta\rho\sigma} h_{\mu\zeta} \partial_\xi \left( \phi \partial_\rho g_{\sigma\nu} \right) \nonumber \\
&\, + \alpha \left( \phi^2 \right) \tilde \epsilon^{\xi\zeta\rho\sigma} g_{\mu\zeta} \partial_\xi \left( \phi \partial_\rho h_{\sigma\nu} \right)
+ \alpha \left( \phi^2 \right) \tilde \epsilon^{\xi\zeta\rho\sigma} h_{\nu\zeta} \partial_\xi \left( \phi \partial_\rho g_{\sigma\mu} \right) 
+ \alpha \left( \phi^2 \right) \tilde \epsilon^{\xi\zeta\rho\sigma} g_{\nu\zeta} \partial_\xi \left( \phi \partial_\rho h_{\sigma\mu} \right) \, .
\end{align}
Let us now choose a condition to fix the gauge as follows
\begin{align}
\label{gfc}
0={\tilde\nabla}^\mu h_{\mu\nu}\, .
\end{align}
Because we are interested in the massless spin-two mode, we also impose the following condition,
\begin{align}
\label{ce}
0=g^{\mu\nu} h_{\mu\nu} \, .
\end{align}
By choosing the conditions in Eqs.~(\ref{gfc}) and (\ref{ce}),
we can reduce Eq.~(\ref{gb4bD4B0}) as follows,
\begin{align}
\label{gb4bD4B}
0=&\, \left[ \frac{1}{2} {\tilde R} + \frac{1}{2} \left\{
 - \frac{1}{8} \omega\left( \phi^2 \right) \phi^2 \partial_\mu \phi \partial^\mu \phi 
 - V\left( \phi^2 \right) \right\} \right] h_{\mu\nu}
 - \frac{1}{16} g_{\mu\nu} \omega\left( \phi^2 \right) \phi^2 \partial^\tau \phi \partial^\eta \phi h_{\tau\eta} 
\nonumber \\
&\, - \frac{1}{2} g_{\mu\nu} h_{\rho\sigma} {\tilde R}^{\rho\sigma} 
 - \frac{1}{2} \left\{ - {\tilde\nabla}^2 h_{\mu\nu} - 2{\tilde R}^{\lambda\ \rho}_{\ \nu\ \mu}h_{\lambda\rho}
+ {\tilde R}^\rho_{\ \mu}h_{\rho\nu} + {\tilde R}^\rho_{\ \mu}h_{\rho\nu} \right\} \nonumber \\
&\, + \alpha \left( \phi^2 \right) \phi \tilde\epsilon^{\zeta\xi\rho\sigma} \partial_\zeta h_{\xi\mu} \partial_\rho g_{\sigma\nu} 
+ \alpha \left( \phi^2 \right) \phi \tilde\epsilon^{\zeta\xi\rho\sigma} \partial_\zeta g_{\xi\mu} \partial_\rho h_{\sigma\nu} \nonumber \\
&\, + \tilde \epsilon^{\xi\zeta\rho\sigma} h_{\mu\zeta} \partial_\xi \left( \alpha \left( \phi^2 \right) \phi \partial_\rho g_{\sigma\nu} \right) 
+ \tilde \epsilon^{\xi\zeta\rho\sigma} g_{\mu\zeta} \partial_\xi \left( \alpha \left( \phi^2 \right) \phi \partial_\rho h_{\sigma\nu} \right) \nonumber \\
&\, + \tilde \epsilon^{\xi\zeta\rho\sigma} h_{\nu\zeta} \partial_\xi \left( \alpha \left( \phi^2 \right) \phi \partial_\rho g_{\sigma\mu} \right) 
+ \tilde \epsilon^{\xi\zeta\rho\sigma} g_{\nu\zeta} \partial_\xi \left( \alpha \left( \phi^2 \right) \phi \partial_\rho h_{\sigma\mu} \right) \, .
\end{align}
The above expression is used when we discuss the propagation of the gravitational wave, which is massless and has spin two. 

We now consider the solutions in the FLRW spacetime with a flat spatial part, 
\begin{equation}
\label{FLRWmetric}
ds^2 = - dt^2 + a(t)^2 \sum_{i=1,2,3} \left( dx^i \right)^2 \, ,
\end{equation}
We shall assume that the axion field $\phi$ depends solely on the cosmic time $t$. 
We also assume $h_{tt}=h_{ti}=h_{it}=0$ and therefore $\sum_{i=1,2,3} h_{ii}=0$. 
The connections and curvatures in the flat FLRW spacetime (\ref{FLRWmetric}) are given by 
\begin{align}
\label{curvatures2}
& \Gamma^t_{ij}=a^2 H \delta_{ij}\ ,\quad \Gamma^i_{jt}=\Gamma^i_{tj}=H\delta^i_{\ j}\, ,\nonumber \\
& {\tilde R}_{itjt}=-\left(\dot H + H^2\right)a^2\delta_{ij}\, ,\quad 
{\tilde R}_{ijkl}= a^4 H^2 \left(\delta_{ik} \delta_{lj} - \delta_{il} \delta_{kj}\right)\, ,\nonumber \\
& {\tilde R}_{tt}=-3\left(\dot H + H^2\right)\ ,\quad {\tilde R}_{ij}= a^2 \left(\dot H + 3H^2\right) \delta_{ij}\ ,\nn
& {\tilde R}= 6\dot H + 12 H^2\, , \quad \mbox{other components}=0\, .
\end{align}
Then one finds 
\begin{align}
\label{deAl}
{\tilde\nabla}^2 h_{tt} =&\, - \tilde\nabla_t \tilde\nabla_t h_{tt} + a^{-2} \delta^{ij} \tilde\nabla_i \tilde\nabla_j h_{tt} 
= - a^{-2} H \delta^{ij} \tilde\nabla_j h_{it} = a^{-2} H^2 \delta^{ij} h_{ij} = 0 \, ,
\nonumber \\
{\tilde\nabla}^2 h_{ij} =&\, - \partial_t^2 h_{ij} + H \partial_t h_{ij} + 2 \dot H h_{ij} + 4 H^2 h_{ij} + a^{-2} \delta^{kl} \partial_k \partial_l h_{ij} \, ,
\end{align}
and $(t,t)$ component of (\ref{gb4bD4B}) vanishes identicaly and $(i,j)$ component has the following form, 
\begin{align}
\label{gb4bD4B2}
0=&\, \frac{1}{2} \left( - \partial_t^2 + H \partial_t + 6 \dot H + 2 H^2 + a^{-2} \delta^{kl} \partial_k \partial_l \right) h_{ij} \nonumber \\
&\, - a^{-1} \epsilon^{klm} \left( 2 \alpha \left( \phi^2 \right) \phi H + \left( 2\alpha' \left( \phi^2 \right) \phi + \alpha \left( \phi^2 \right) \right) \dot\phi \right) 
\left( \delta_{lj} \partial_m h_{ki} + \delta_{li} \partial_m h_{kj} \right) \, . 
\end{align}
Here we have used (\ref{Fb}). 

In order to make the situation clear, we now consider the gravitational wave propagating along the $z$-direction, that is,
$h_{ij} \propto \e^{-i\left( \omega t - k(t) z \right)}$, with an angular frequency $\omega$ and a wavenumber $k$. 
As the spacetime has a translational invariance under the shifts of $x$, $y$, and $z$ directions, 
one may assume $k$ does not depend on the spatial coordinates $x$, $y$, and $z$. 
We now consider that the frequency and wavenumber are much larger than the time-evolution of the universe, that is, 
$\omega^2, \, k(t)^2 \gg H^2,\, \left| \dot H \right|$. 
Under the assumption, we neglect the derivatives of $k(t)$ and the amplitudes with respect to the time $t$. 

Because we assume $h_{tt}=h_{ti}=h_{it}=0$ and therefore $\sum_{i=1,2,3} h_{ii}=0$, one finds 
\begin{align}
\label{CGRa2}
h_{iz}=0\, , \quad h_{xx}=- h_{yy} = h_+ \e^{-i\left(\omega (t) t - k z \right)} \, , \quad
h_{xy}=h_{yx} = h_\times \e^{-i\left(\omega (t) t - k z \right)} \, ,
\end{align}
with complex constants $h_+$ and $h_\times$, which express the polarisations of the gravitational wave. 
Then the $(x,x)$ and $(y,y)$ components in (\ref{gb4bD4B2}) give the following identical results, 
\begin{align}
\label{xxyy}
0 =&\, \frac{1}{2} \left( \omega^2 + i H \omega + 6 \dot H + 2 H^2 + a^{-2} k(t)^2 \right) h_+(t) \nonumber \\
&\, + ik a^{-1} \left( 2 \alpha \left( \phi^2 \right) \phi H + \left( 2\alpha' \left( \phi^2 \right) \phi + \alpha \left( \phi^2 \right) \right) \dot\phi \right) h_\times (t)\, .
\end{align}
On the other hand, $(x,y)$ or $(y,x)$ component in (\ref{gb4bD4B2}) gives 
\begin{align}
\label{xy}
0 =&\, \frac{1}{2} \left( \omega^2 + i H \omega + 6 \dot H + 2 H^2 + a^{-2} k(t)^2 \right) h_\times(t) \nonumber \\
&\, - ik a^{-1}\left( 2 \alpha \left( \phi^2 \right) \phi H + \left( 2\alpha' \left( \phi^2 \right) \phi + \alpha \left( \phi^2 \right) \right) \dot\phi \right) h_+ (t)\, .
\end{align}
In order that Eqs.~(\ref{xxyy}) and (\ref{xy}) have non-trivial solutions for $h_+$ and $h_\times$, we obtain, 
\begin{align}
\label{disp}
& \frac{1}{2} \left( \omega^2 + i H \omega + 6 \dot H + 2 H^2 + a^{-2} k(t)^2 \right) h_\times(t) \nonumber \\
& = \pm k a^{-1}\left( 2 \alpha \left( \phi^2 \right) \phi H + \left( 2\alpha' \left( \phi^2 \right) \phi + \alpha \left( \phi^2 \right) \right) \dot\phi \right) \, .
\end{align}
Eq.~(\ref{disp}) gives the dispersion relation. 
The signature $\pm$ in the r.h.s. of Eq.~(\ref{disp}) corresponds to the polarizations, that is, left-handed and right-handed modes. 
Therefore, the left-handed mode has a dispersion relation different from that of the right-handed mode. 
When we solve Eq.~(\ref{disp}) with respect to $\omega$, the solution $\omega$ becomes a complex number function of $t$. 
The imaginary part expresses the enhancement or decrease of the amplitude of the gravitational wave. 

\section{Axion field misalignment and axion particle production}\label{SecVII}

Due to the constraints of parity, the potential must be an even function of the axion field and therefore, if the minimum could be the origin of the potential 
as in the QCD axion, it is difficult to generate the misalignment. 
Since the coupling of the pseudo-scalar quantity $R_\mathcal{Q}$ in (\ref{Qps}) with the axion field must be an odd function of the axion field due to the parity symmetry, 
if the pseudo-scalar quantity $R_\mathcal{Q}$ has a non-trivial value due to fluctuations or something else in the early universe, the minimum is shifted to a non-trivial value 
and the misalignment of the axion field occurs. 
If the expectation value of the pseudo-scalar field vanishes after that, the axion field starts oscillating around the minimum of the potential and the axion particles are generated. 

In order that $R_\mathcal{Q}$ has a non-trivial value, the universe cannot be the FLRW universe where the spatial part is isotropic and inhomogeneous. 
When the symmetry breaking occurs, there appear the bubbles of the true vacua and the universe cannot be isotropic or inhomogeneous. 
Furthermore, the axion production must occur after inflation, so that the density does not become too small due to the rapid expansion of the universe. 
Therefore, such a period when the universe is neither isotropic nor inhomogeneous could be the period of the electroweak phase transition. 
The energy scale of the electroweak phase transition is about 160\, $\mathrm{GeV}$, the energy density of the matter should 
be $\left(160\, \mathrm{GeV}\right)^4\sim 10^{9}\, \mathrm{GeV}^4$. 
Then the nonmetricity scalar $Q$ could be estimated as $Q\sim10^{9}\, \mathrm{GeV}^4/\left( 10^{19}\, \mathrm{GeV}\right)^2 = 10^{-29} \, \mathrm{GeV}^2$. 
Because the mass dimension of $R_\mathcal{Q}$ is identical with that of $Q$, we may also estimate $R_\mathcal{Q}$ as $R_\mathcal{Q}\sim 10^{-29} \, \mathrm{GeV}^2$. 

In order to make the story definite, we choose 
\begin{align}
\label{choice}
\omega \left(\phi^2 \right) = \frac{4}{\phi^2} \, , \quad V \left(\phi^2 \right) = \frac{1}{2} {m_\phi}^2 \phi^2 \, , \quad 
\alpha \left(\phi^2 \right) = \alpha_0\, .
\end{align}
Then the axion becomes canonical, the mass is given by $m_\phi$, and $\alpha \left(\phi^2 \right)$ becomes a constant $\alpha_0$. 

When $R_\mathcal{Q}$ does not vanish, the effective potential in (\ref{Veff}) has the following form, 
\begin{align}
\label{VeffB}
V_\mathrm{eff} (\phi) = \frac{1}{2} {m_\phi}^2 \phi^2 + \alpha_ 0 \phi R_\mathcal{Q}\, ,
\end{align}
whose minimum is given by 
\begin{align}
\label{phimin}
\phi_\mathrm{min} = - \frac{\alpha_ 0 R_\mathcal{Q}}{{m_\phi}^2}\, .
\end{align}
Then, the misalignment of the axion field occurs. 
After the electroweak phase transition, $R_\mathcal{Q}$ vanishes again but shifted axion field in (\ref{phimin}) gives non-trivial 
potential energy
\begin{align}
\label{Vmin}
V\left( \phi_\mathrm{min} \right) = \frac{{\alpha_ 0}^2 {R_\mathcal{Q}}^2}{{2 m_\phi}^2} \, .
\end{align}
Then the axion field oscillates with an angular frequency $m_\phi$ and the initial amplitude is given by $\left| \phi_\mathrm{min}\right|$ in (\ref{phimin}). 
The Hubble rate $H$ plays the role of the resistance for the motion as usual. 
Furthermore, the oscillation produces the axion particles, which are quanta of the axion field. 
Then the amplitude of the oscillation is decreased and the potential energy in (\ref{Vmin}) is converted to the energy density of the axion particles. 
Therefore, the energy density of the axion particles is almost equal to or a little less than (\ref{Vmin}). 
The number density $n_\mathrm{a}$ of the axion particles can be estimated by dividing the energy density by the mass $m_\phi$ of the axion. 
\begin{align}
\label{numberdensity}
n_\mathrm{a} \sim \frac{{\alpha_ 0}^2 {R_\mathcal{Q}}^2}{{2 m_\phi}^3} \, .
\end{align}
Since $R_\mathcal{Q}\sim 10^{-29} \, \mathrm{GeV}^2$, by adjusting the parameters $\alpha_ 0$ and $m_\phi$, we may realise the realistic dark matter density. 

In (\ref{choice}), we have chosen that $V(\phi)=0$ when $\phi=0$. 
If the minimum of $V(\phi)$ is positive at the minimum, the minimum value takes the role of the cosmological constant, which may generate the late-time 
accelerating expansion of the universe. 

Usually, the axion misalignment is believed to have occurred during the inflationary era because the axion decay constant could be estimated to 
typically ranges from $10^9 - 10^{12}\, \mathrm{GeV}$ if the origin of the axion is QCD. 
In our model, the axion does not always originate from QCD; therefore, the axion could be more accurately referred to as so-called axion-like particles (ALPs). 
Therefore, the energy scale of the misalignment in our model is not constrained as in the QCD axion. 
Furthermore, the misalignment associated with the electroweak phase transition might relax the $S_8$ problem as we discuss in the next section. 

\section{$S_8$ problem}\label{SecVIII}

Recent DESI observation \cite{DESI:2024mwx} seems to indicate a transition from $w<-1$ to $w>-1$ in the dark energy. 
A recent proposal to solve this problem is that, instead of considering the transition from $w<-1$ to $w>-1$, which we may call ``inverse phantom crossing'', 
the modification of the dark matter sector was considered \cite{Khoury:2025txd}. 
In the scenario, the dark matter decreases more slowly than $1/a^3$, which is usually predicted from the energy conservation of the dust matter ($w=0$). 
As we are considering the total energy density, the DESI observation seems to indicate that there might have occurred the inverse phantom crossing 
of the dark energy sector only in the case that we assume the usual $1/a^3$ behaviour of the dark matter. 

The behaviour of the dark matter, which decreases more slowly than $1/a^3$, could be realised if we consider more slow-roll potential than that given in (\ref{choice}) 
for the axion as in the inflaton model in the context of this paper. 
Then, after the misalignment of the axion field, the axion field goes down slowly, which looks like a slower decrease than $1/a^3$. 
This solves the $S_8$ problem. 

We now propose the following potential instead of the potential in (\ref{choice}), 
\begin{align}
\label{slpot2}
V \left(\phi^2\right) = \frac{1}{\beta^2}\left( 1 - \e^{-\frac{{m_\phi}^2 \beta^2}{2} \phi^2} + \epsilon \right) \, .
\end{align}
Here $\beta$ and $\epsilon$ are constant, and we assume $\epsilon$ is positive. 
Furthermore, $\frac{\epsilon}{\beta^2}$ plays the role of a small cosmological constant. 
When $\phi\sim 0$, the potential $V \left(\phi^2\right)$ in (\ref{slpot2}), we find $V\left( \phi^2 \right) \sim \frac{1}{2} {m_\phi}^2 \phi^2 + \frac{\epsilon}{\beta^2}$, 
that is, the potential in (\ref{choice}) with the small cosmological constant $\frac{\epsilon}{\beta^2}$ is obtained. 
On the other hand, when $\left| \phi \right|$ is large enough, $\phi^2 \gg \frac{1}{{m_\phi}^2 \beta^2}$, $V \left(\phi^2\right)$ goes to a constant, 
$V \left(\phi^2\right) \to \frac{1}{\beta^2}\left( 1 + \epsilon \right) $, and therefore it behaves like a large cosmological constant. 
This tells that the energy density of the axion field with an particles could decrease much more slowly than $1/a^3$. 

Instead of the $\alpha \left( \phi^2 \right)$ in (\ref{choice}), we choose 
\begin{align}
\label{slalpha}
\alpha \left(\phi^2\right) = \frac{\alpha_1}{\beta^2} \e^{-\frac{{m_\phi}^2 \beta^2}{2} \phi^2} \, .
\end{align}
Here $\alpha_1$ is a constant. 
The reason why we choose (\ref{slalpha}) is because if we choose $\alpha \left( \phi^2 \right)$ as in (\ref{choice}), 
the effective potential in (\ref{Veff}) does not have any minimum when $R_\mathcal{Q}$ is large. 

The minimum of the effective potential (\ref{Veff}) with (\ref{slpot2}) and (\ref{slalpha}) is found by 
\begin{align}
\label{slminimum}
0 = \frac{d V_\mathrm{eff}\left( \phi^2 \right)}{d\phi} = {m_\phi}^2 \e^{-\frac{{m_\phi}^2 \beta^2}{2} \phi^2} \phi 
+ \frac{\alpha_1}{\beta^2} \e^{-\frac{{m_\phi}^2 \beta^2}{2} \phi^2} \left( 1 -{m_\phi}^2 \beta^2 \phi^2 \right) R_\mathcal{Q}\, ,
\end{align}
whose solution is given by 
\begin{align}
\label{slsol}
\phi = \phi^\pm \equiv \frac{{m_\phi}^2 \pm \sqrt{ {m_\phi}^4 + \frac{4 {\alpha_1}^2 {m_\phi}^2}{\beta^2} {R_\mathcal{Q}}^2 }}{2\alpha_1 {m_\phi}^2R_\mathcal{Q}} 
= \frac{1 \pm \sqrt{ 1 + \frac{4 {\alpha_1}^2 {R_\mathcal{Q}}^2}{\beta^2 {m_\phi}^2}}}{2\alpha_1 R_\mathcal{Q}} \, .
\end{align}
In $\pm$ of (\ref{slsol}), the plus signature ``$+$'' corresponds to the maximum of the effective potential $V_\mathrm{eff}\left( \phi^2 \right)$ 
and the minus signature ``$-$'' to the minimum. 
Therefore, by the phase transition, the axion field is shifted to $\phi^-$. 
If the absolute value of $\alpha_1$ is large enough, after the phase transition, the axion is shifted to the place where the gradient of 
the potential $V\left( \phi^2 \right)$ is small, and the axion field moves to the origin very slowly, 
which makes the energy density of the axion field with axion particles decrease much slower than $1/a^3$. 

In the above scenario, the period where the axion field has a non-trivial value becomes longer due to the slow-roll. 
Then the gravitational wave propagates in the period, there appear differences in the dispersion relations of the left-handed mode 
and the right-handed mode, which might be found by future observations. 

\section{Beginning of Inflation}\label{SecIX}

Usually, we discuss the end of inflation, but not so much the beginning. 
In our proposal in this paper, due to the pseudo-scalar quantity, the misalignment is generated by the phase transitions associated with the symmetry breaking. 
In the previous sections, we considered the electroweak phase transition after inflation so as not to break the solution of the monopole problem. 
The misalignment could, however, be generated by the GUT phase transition. 
If the axion is misaligned by the transition, as in the case of the electroweak phase transition, 
and if the potential is slow-roll type in scale as in (\ref{slpot2}), the axion plays the role of the inflaton. 
That is, the inflation could begin by the GUT phase transition, which is, of course, natural to solve the monopole problem. 

We may use the identical potential $V\left( \phi^2 \right)$ in (\ref{slpot2}) and 
the identical coefficient function $\alpha\left( \phi^2 \right)$ in (\ref{slalpha}), again. 
Then the solution of the minimum of the effective potential (\ref{Veff}) is, again, given by (\ref{slsol}). 
The difference is the magnitude of $R_\mathcal{Q}$. 
As the scale is given by the symmetry breaking of the gauge symmetry of GUT, we can estimate $R_\mathcal{Q}$ as 
$R_\mathcal{Q} \sim \left( 10^{16}\, \mathrm{GeV} \right)^2$. 
Therefore, the inflation could start from the flatter part of the potential compared to the axion dark matter production in the last section. 

In this scenario, during the inflation, the axion field has a non-trivial value and therefore there appear
the differences in the dispersion relations of the left-handed mode and the right-handed mode of the gravitational wave. 
These differences may affect the structure formation. 
As long as we consider the model where $\alpha\left( \phi^2 \right)$ is given by (\ref{slalpha}), $\alpha\left( \phi^2 \right)$ could be very small 
due to the Gaussian factor $\e^{-\frac{{m_\phi}^2 \beta^2}{2} \phi^2}$, and therefore the difference in the polarisation could be suppressed. 
Of course, this situation depends on the details of the model. 

At the end of inflation, the matter could be generated by the oscillation of the axion field. 
Still, we may expect that the matter field could not be generated so much, and the axion particles could be mainly produced. 
A way of realising this situation is that the coupling between the matter and the axion depends on $Q$. 
An example could be given by the replacement, 
\begin{align}
\label{matteraxion}
\mathcal{L}_\mathrm{matter} \to \left( 1 + \gamma \e^{\frac{Q^2}{{Q_0}^2}} \phi^2 \right) \mathcal{L}_\mathrm{matter} \, .
\end{align}
Here $\mathcal{L}_\mathrm{matter} $ is the Lagrangian density of the matter, and $\gamma$ and $Q_0$ are constants. 
After the inflation, the non-metricity scalar $Q$ could be rather large, and the coupling between the matter and the axion field is large. 
Therefore, the matter could be mainly produced by the oscillations of the axion field. 
On the other hand, after the electroweak phase transition, $Q$ could be rather small and the coupling 
between the matter and the axion field could not be so large, and therefore, the axion particles are mainly produced. 

\section{Summary and conclusion}\label{SecXI}

In this paper, we proposed scenarios which may solve the problems of the axion misalignment, the $S_8$ problem, and the beginning of inflation 
in the framework of the non-metricity gravity coupled with an axion field. 
For this purpose, we introduced the pseudo-scalar quantity $R_\mathcal{Q}$ in (\ref{Qps}), which is an analogue of the Chern-Simons invariant. 

The coupling between $R_\mathcal{Q}$ in (\ref{Qps}) and the axion field $\phi$ does not play any role when we consider the evolution of the 
FLRW universe, where the spatial part is homogeneous and isotropic. 
The propagation of the gravitational wave is modified by the coupling, and the dispersion relation of the left-handed mode 
is different from that of the right-handed mode. 

When the phase transition associated with the gauge symmetries of the electroweak symmetry and GUT occurs, 
the pseudo-scalar quantity $R_\mathcal{Q}$ has a non-trivial value, which generates the misalignment of the axion field and produces 
the axion particles, which may be a candidate for the dark matter. 

By considering the potential whose gradient is small, we gave the scenarios which may explain the observation of DESI 2025
and solve the $S_8$ problem. 
This mechanism may also start the inflation by the GUT phase transition and finally includes the reheating, where matters are generated. 

During the period when the axion field has a non-trivial value, the parity symmetry is broken, and the propagation of the gravitational wave and the structure formation are affected. 
Although we have not found the parity violation \cite{Cahn:2021ltp}, any small violation of the parity might be found, which may give evidence towards the axion cosmology. 

We now summarise the timeline of the scenario in this paper. 
First, due to the GUT phase transition, the axion misalignment occurs, which may generate inflation. 
The inflation ends when the axion goes to the bottom of the potential, and due to the oscillation of the axion field, matter particles 
could be mainly produced. 
After that, the axion misalignment could occur again due to the electroweak phase transition. 
Due to this misalignment, the $S_8$ problem might be relaxed, and the axion particles could be produced to be dark matter. 


Similar scenarios could be realised by using the Chern-Simons invariant 
$R_\mathrm{CS} \equiv \tilde\epsilon^{\mu\nu\rho\sigma} R^\tau_{\ \lambda\mu\nu}R^\lambda_{\ \tau\rho\sigma}$ 
instead of the pseudo-scalar quantity $R_\mathcal{Q}$ in (\ref{Qps}) and using $R$ instead of $Q$. 
The problem is the scale of the term coupled with the axion field $\phi$. 
In the model (\ref{model1}), in order that the term $\alpha\left( \phi^2 \right) \phi R_\mathcal{Q}$ could be compatible with the first term $Q$ and the term gives 
non-trivial contributions, the order of $\alpha\left( \phi^2 \right)$ should be always unity, $\alpha\left( \phi^2 \right) \sim \mathcal{O}(1)$ 
because $Q\sim R_\mathcal{Q}$. 
Note that we have chosen the mass dimension of the axion field to vanish, and therefore we may assume $\phi \sim \mathcal{O}(1)$. 
As is discussed in Section~\ref{SecVII}, we estimated $R_\mathcal{Q}$ as $R_\mathcal{Q}\sim Q \sim 10^{-29} \, \mathrm{GeV}^2$ during the electroweak phase transition. 
Because the Chern-Simons invariant $R_\mathrm{CS}$ is given by the square of the curvatures, we may estimate 
$R_\mathrm{CS} \sim 10^{-58} \, \mathrm{GeV}^4$, which is very small. 
Then in order that the term like $\alpha_\mathrm{CS} \phi R_\mathcal{CS}$, instead of $\alpha\left( \phi^2 \right) \phi R_\mathcal{Q}$, 
gives a non-trivial contribution, or this term has a magnitude comparable to 
the scalar curvature $R \sim 10^{-29} \, \mathrm{GeV}^2$, the coefficient $\alpha_\mathrm{CS}$ must be unnaturally large, 
$\alpha_\mathrm{CS} \sim 10^{29} \, \mathrm{GeV}^{-2}$. 
If we choose $\alpha_\mathrm{CS}$ so large, the term $\alpha_\mathrm{CS} \phi R_\mathcal{CS}$ becomes too large during the phase transition of the GUT 
and it becomes difficult to realise the inflation consistently, not as in Section~\ref{SecIX}. 


We may consider the generalisation of the model (\ref{model1}). 
A simple generalization of (\ref{model1}) is given by (\ref{model2}). 
In the theory (\ref{model2}), the axion coupling is not changed from that in (\ref{model1}). 
Therefore, the qualitative structure of the axion particle production and behavior of the axion field and the axion particles are not really changed. 

When we consider the gravitational waves, the only propagating mode in the flat background of $f(Q)$ gravity is given by the standard gravitational wave, which is 
massless and has spin two \cite{Capozziello:2024vix, Capozziello:2024jir}. 
Therefore, the situation is not really changed even in the model (\ref{model2}), although the parity-odd wave could appear, which is associated with the axion. 

As the difference between the scalar curvature $\tilde R$ and the non-metricity scalar $Q$ is a total derivative, 
$\tilde R = Q - {\tilde\nabla}_\alpha\left(Q^{\alpha}-\tilde{Q}^{\alpha}\right)$, 
the models including the difference $B=-C=Q-\tilde R$ have been proposed \cite{Capozziello:2023vne, Gadbail:2023mvu}. 
Then more generalisations of theory (\ref{model2}) could be proposed by 
\begin{align}
\label{model3}
S= \frac{1}{2\kappa^2} \int d^4 x \sqrt{-g} \left\{ f\left(Q, B, \phi^2 \right) - \frac{1}{8} \omega\left( \phi^2 \right) \phi^2 \partial_\mu \phi \partial^\mu \phi 
 - \alpha\left( Q, B, \phi^2 \right) \phi R_\mathcal{Q} \right\} \, .
\end{align}
Here, the potential of the axion is included in the first term $f\left(Q, B, \phi^2 \right)$. 
Then the axion coupling also depends on $Q$ and $B$, which may control the axion particle production. 

In $f(Q, B)$ gravity, a scalar mode corresponding to the metric scale or the scalar curvature itself appears. 
The mode propagates with a non-trivial mass \cite{Capozziello:2024zij}. 
Then, even in the model (\ref{model3}), there could appear the scalar mode. 
As the scalar mode is scalar with parity even, this mode could not be mixed with the axion field with parity odd. 

We should stress that in order to include the coupling with the pseudo-scalar quantity like $R_\mathcal{Q}$, the Chern-Simons invariant 
$R_\mathcal{CS} \equiv \tilde\epsilon^{\mu\nu\rho\sigma} g^{\xi\zeta} g^{\eta\tau} \tilde R_{\mu\nu\xi\eta} \tilde R_{\rho\sigma\zeta\tau}$, 
or the instanton density $F\tilde F$, we need a pseudo-scalar field like the axion. 
For example, if we consider the following action including a field $\chi$ instead of (\ref{model1}), 
\begin{align}
\label{inconsistentmodel}
S= \frac{1}{2\kappa^2} \int d^4 x \sqrt{-g} \left\{ Q - \frac{1}{2} \partial_\mu \chi \partial^\mu \chi 
 - V_\chi\left( \chi \right) - \alpha_\chi \left( \chi \right) R_\mathcal{Q} \right\} \, ,
\end{align}
we cannot assign any parity on the field $\chi$, and $\chi$ cannot be a scalar field nor a pseudo-scalar field. 
Therefore, in the action (\ref{inconsistentmodel}), the parity symmetry is explicitly broken, and the breaking is generally large if 
there is no parameter which controls the magnitude of the breaking. 
Therefore, this kind of model could be inconsistent with the recent observation \cite{Cahn:2021ltp}.

\section*{ACKNOWLEDGEMENTS}

This work was partially supported by the program Unidad de Excelencia Maria de Maeztu CEX2020-001058-M, Spain (S.D.O).

\end{document}